# Observation of Optical Gain in Er-Doped GaN Epilayers

V. X. Ho,[1] Y. Wang,[1] B. Ryan,[1] L. Patrick,[1] H. X. Jiang,[2] J. Y. Lin,[2] N. Q. Vinh[1,*]

[1]Department of Physics & Center for Soft Matter and Biological Physics, Virginia Tech, Blacksburg, Virginia 24061, USA
[2]Department of Electrical and Computer Engineering, Texas Tech University, Lubbock, Texas 79409, USA

*Corresponding author: vinh@vt.edu

**Abstract**

Rare-earth based lasing action in GaN semiconductor at the telecommunication wavelength of 1.5 µm has been demonstrated at room temperature. We have reported the stimulated emission under the above bandgap excitation from Er doped GaN epilayers prepared by metal-organic chemical vapor deposition. Using the variable stripe technique, the observation of the stimulated emission has been demonstrated through characteristic features of threshold behavior of emission intensity as functions of pump intensity, excitation length, and spectral linewidth narrowing. Using the variable stripe setup, the optical gain up to 75 cm$^{-1}$ has been obtained in the GaN:Er epilayers. The near infrared lasing from GaN semiconductor opens up new possibilities for extended functionalities and integration capabilities for optoelectronic devices.

**I. INTRODUCTION**

Near infrared semiconductor lasers are essential for optical telecommunication, industrial processing, military defense, medical applications, and spectroscopy and imaging.[1-4] These semiconductor light sources can be readily integrated with the silicon microelectronics technology and/or optical fiber-based devices. In this context rare-earth (RE) doping of semiconductors is of particular importance for applications in optoelectronics.[4-6] Among various RE doped semiconductors for near-infrared lasers, erbium (Er) providing light emission at the 1.5 µm window via the $^4I_{13/2} \rightarrow {}^4I_{15/2}$ radiative transition becomes a prominent candidate for near infrared applications.[1, 6] This technologically important wavelength coincides with the minimum absorption band of optical fibers currently used in telecommunication, and in the eye-safe wavelength region.[6, 7] In the past, the goal of obtaining 1.5 µm Er-doped semiconductor lasers was unsuccessfully attempted using crystalline Si,[6, 8-10] latter of SiO$_2$:Er sensitized with Si nanocrystals,[11-14] or GaAs and AlGaAs because of the strong thermal quenching effect occurring in narrow and/or indirect bandgap materials.[15, 16]

Solid-state lasers based on RE ions doped transparent solid materials including oxide, fluorides, ceramic, and glasses provide high power light sources, sharp, atomic-like, and temperature independent luminescence bands at room temperature (RT).[2, 5] With a large bandgap, a common way to excite Er ions in such host materials has been done using optical pumping sources in which photon energies match higher-lying intra-4*f* transitions. This resonant pump method provides an excitation cross-section of about three to five orders of magnitude lower compared to those of the above bandgap excitation.[7] This limitation is a big obstruction to further progress towards RT 1.5-µm lasers based on Er in micro-optoelectronic applications.

GaN semiconductor is an attractive host material for RE doping because of the direct and wide (~3.3 eV) bandgap properties. Er-doped GaN material exhibits a significant low degree of thermal luminescent quenching, and strong emission at RT under either electrical or optical excitation.[7, 15, 17] We have successfully prepared Er-doped GaN epilayer on Si (001) and sapphire substrates by metal organic chemical vapor deposition (MOCVD) with good crystallinity and high percentage of $Er^{3+}$ optical centers, resulting in predominant photoluminescence (PL) at RT.[18-20] These prior works have provided necessary base to achieve near-infrared RT lasers based on Er-doped GaN (GaN:Er) semiconductor material. Recently, we have reported a lasing action in GaN/AlN multiple quantum wells in the infrared 1.5 μm region.[17] The lasing in this structure is only obtained in a small volume.

In this paper, we describe techniques and methods to obtain lasing from GaN:Er epilayers which can be grown in a large volume (bulk GaN material). GaN has a high thermal conductivity ($\kappa \approx 253$ W/m.K)[21], and a low thermal expansion coefficient ($\alpha \approx 3.2 \times 10^{-6}$ $K^{-1}$)[22-24] compared with other semiconductors and ceramic including silicon ($\kappa \approx 130$ W/m.K, $\alpha \approx 2.6 \times 10^{-6}$ $K^{-1}$),[24, 25] GaAs ($\kappa \approx 50$ W/m.K, $\alpha \approx 6.5 \times 10^{-6}$ $K^{-1}$)[25], and YAG ($\kappa \approx 12$ W/m.K, $\alpha \approx 6.4 \times 10^{-6}$ $K^{-1}$)[26, 27]. Thus, we can obtain high energy and high power lasers in the GaN material. Here, we report the RT lasing in GaN:Er epilayers at 1.5 μm by the variable stripe length technique, and an optical gain up to 75 $cm^{-1}$ under the above bandgap excitation.

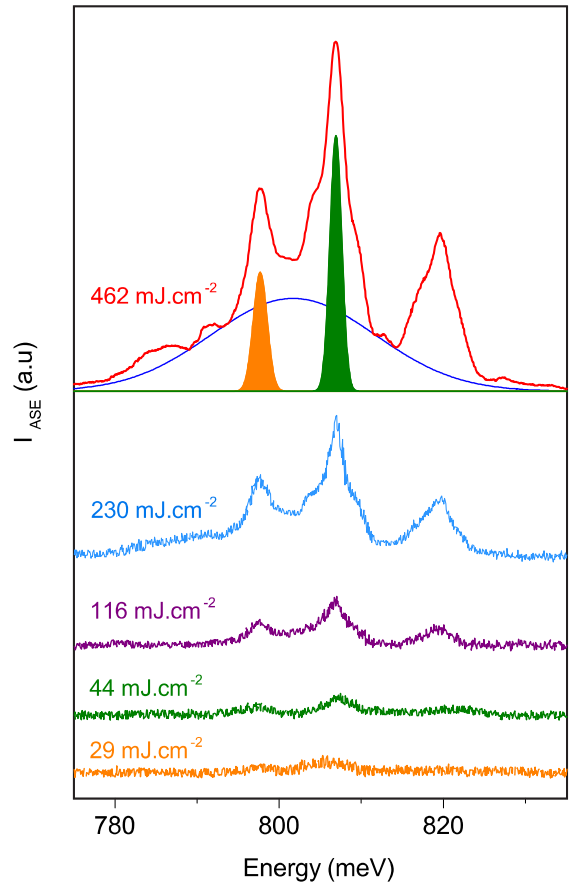

**Figure 1.** The edge PL spectra from a GaN:Er epilayer using a UV Ar laser, $\lambda_{exc}$ = 351 nm, for the above bandgap excitation. The excitation length was 0.2 mm and the pump fluence is varied from 0.5 to 500 $mJ.cm^{-2}$. The PL signal at low pump intensity shows a broad spectrum with a typical FWHM of ~22 meV. When the pump intensity from the UV laser is high enough to have a net gain in the optical layer, the spontaneous photons are exponentially amplified. Under this condition, spectral linewidths become narrower. The spectrum at high pump intensity was deconvolved into several Gaussian peaks. The most prominent PL lines at 797.71 and 806.89 meV show a FWHM of 1.60 ± 0.35 meV.

## II. RESULTS AND DISCUSSION

The Er-doped GaN materials were grown on a (0001) c-plane sapphire substrate in a horizontal reactor MOCVD system.[19] Prior to the growth GaN:Er epilayer, a 20-nm low temperature GaN layer as a buffer, and a 0.2-μm un-doped GaN epilayer were first grown on the sapphire substrate. This was then followed by the growth of the GaN:Er epilayer of 0.5 μm thickness. The temperature for the growth process of Er-doped GaN epilayers was set at 1040 °C. An optimum Er concentration was about ~2 × $10^{20}$ cm$^{-3}$ to obtain the strongest PL intensity. A high degree of crystallinity without a second phase formation in the GaN:Er layers has been confirmed by X-ray diffraction measurements.[28] Details of growth conditions and epilayer structure were described in previous papers.[19, 28] The percentage of Er optical centers has been estimated by comparing the PL intensity at RT between the GaN:Er material and a $SiO_2$:Er reference sample under the resonant excitation, $^4I_{15/2} \rightarrow {}^4I_{9/2}$ transition, providing by a tunable Ti:Sapphire laser at 809 nm. The estimated percentage of $Er^{3+}$ ions emitting 1.5 μm photons is about 68% of the total Er ions.[20] The achievement of the high percentage of active Er ions indicates a high potential for realizing optical amplification in GaN:Er epilayers.

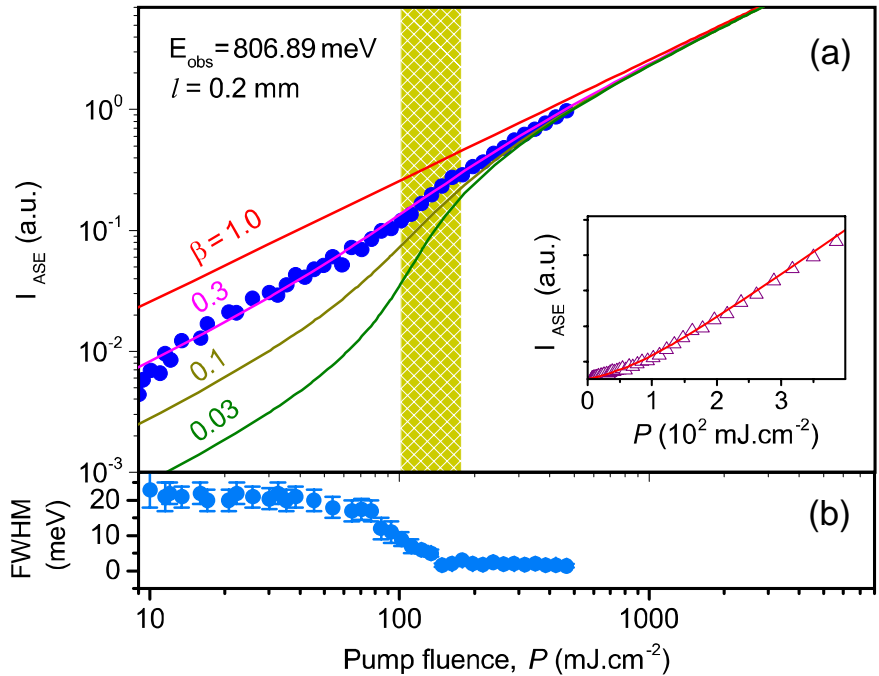

**Figure 2.** Output intensity of the edge-emission for the PL line at 806.89 meV as a function of pump fluence with the excitation length of 0.2 mm. (a) The light-in-light-out data on the log-log scale shows a lasing threshold of 110 mJ.cm$^{-2}$. The result was fitted to the S-curve model. The best fitting curve has been obtained for the coupling factor, $\beta$ = 0.3, for the experimental data. Inset: The edge-emission PL intensity showing a threshold for a transition from the spontaneous emission to the amplified stimulated emission. (b) The FWHM of the PL intensity from the edge of the sample becomes narrower.

To determine the threshold and the net optical gain in the Er-doped GaN epilayer under the above bandgap excitation, we have performed edge-emission measurements using the well-established variable stripe length (VSL) technique.[29] The GaN:Er sample was optically excited by a UV Argon laser providing photons at 351 nm (or 3.531 eV) at RT. The laser beam was expanded by a beam expander, and selected a center part with a 4x4-mm square slit in order to achieve a square homogeneous excitation. After that, the laser beam was focused onto the top surface of the sample using a cylindrical lens with the focal length of 75 mm to obtain a stripe-shaped excitation beam of 8-μm width. The homogeneous stripe-laser beam was verified by scanning the beam profile using a UV photo-detector mounted on a two-dimensional linear stage. An adjustable shield was placed on the top of the sample to control the excitation length up to 1000 μm.[17] To achieve lasing operation, an optical cavity has been created from both polished edges of the sample. Two converging lenses with focal lengths of 10 and 24 cm were used to collect the edge-emission. The emission was focused onto an entrance slit of a high-resolution spectrometer (Horiba iHR550). We obtain the resolution of PL spectra of 0.05 nm. The signal was detected by a sensitivity photodetector (InGaAs DSS-IGA).[30]

Light emitted from the sample surface has spontaneous emission features, whereas light collected from the edge of the sample contains both spontaneous and stimulated emission characteristics. In contrast to the spontaneous emission on the surface, spectral peaks of the PL intensity of the GaN:Er epilayer from the edge of the sample become narrower, and their intensity grows exponentially when pump fluence or excitation length increases (Fig. 1). As previously reported,[19, 30, 31] the surface emission exhibited a broad peak round 1.5 μm with the full-width at half maximum (FWHM) of ~45 nm (~22 meV) under the above bandgap excitation, and the PL intensity as a function of pump power at 1.5 μm showed a linear and saturated behavior. This is typical for spontaneous emission of GaN:Er epilayers at RT. The broadening of PL spectrum at RT is due to the thermal population of higher-lying sublevels of the $^4I_{13/2}$ manifold and multiple optical centers in the GaN:Er material. However, the PL spectra of the edge-emission at high excitation pump fluence show several narrow PL lines (Fig. 1). Typically, the optical gain obtains maximum around the peaks of the spontaneous emission spectrum, thus, the "*gain narrowing*" of PL lines appears in the spectra at high pump power.[32, 33] A deconvolution into several Gaussian components of the PL spectrum has been shown in the Fig. 1. The intensity of narrow PL lines at 797.71 and 806.89 meV exhibits an increasing from linear to exponential behavior.

We have investigated the PL intensity of the edge-emission from Er optical centers by varying the pump intensity of the UV laser from 0.5 to 500 mJ.cm$^{-2}$ under a fixed excitation length of 0.2 mm. Figure 2a shows PL intensity at 806.89 meV as a function of pump fluence. In the linear plot (Fig. 2a, inset) the PL intensity shows a linear dependence at low pump fluence and a super-linear behavior at high pump fluence, indicating a transition from the spontaneous emission to the amplified stimulated emission. Above a certain value of pump fluence, the PL lines become narrower. The observed spontaneous-to-simulated emission transition is a clear evidence of optical amplified spontaneous emission (ASE) in GaN:Er epilayers.

To estimate the gain threshold value, we have employed the coupled rate-equation analysis to explore the experimental results of the light-in-light-out (L-L) curve of Er-doped GaN epilayers with a simple cavity. Figure 2a represents the logarithmic plot of the PL intensity at 806.89 meV as a function of the pump fluence with a fixed excitation length of 0.2 mm. Specifically, the rate-equations describe the evolution of the carrier density, *N*, in the active region, and the photon density, *P*, in the cavity mode under optical pumping:[34]

$$\frac{dN}{dt} = \frac{\eta_a R_p}{\hbar \omega V} - \frac{N}{\tau_{sp}} - \frac{N}{\tau_{nr}} - v_g g P \qquad (1)$$

$$\frac{dP}{dt} = \Gamma v_g (g - g_{th}) P + \Gamma \beta \frac{N}{\tau_{sp}} \qquad (2)$$

where $R_p$ is the optical pumping rate; $\eta_a$ is the fraction of the pump fluence absorbed by the active medium in a volume, $V$, of the GaN:Er epilayer; $\hbar\omega$ is the photon energy of the pump beam; $1/\tau_{sp}$ is the spontaneous emission rate with the decay time constant, $\tau_{sp}$, of 3.3 ± 0.3 ms;[30] $1/\tau_{nr}$ is the non-radiative recombination rate; $v_g = c/n$ is the group velocity; $n$ = 2.6 is the refractive index of the material; $\Gamma$ is the optical confinement factor of the lasing mode; β describes the amount of spontaneous emission that couples into the laser cavity mode; $g$ is the material gain; and $g_{th}$ is the threshold gain. A linear relation between the material gain, $g$, and carrier density, $N$, is assumed in the active region, $g = a(N - N_{tr})$, where $a$ is a material constant, and $N_{tr}$ is the transparency carrier density. The transparency carrier number does not affect the fitting result, we set this value to be zero. Non-radiative decay processes produce additional losses, which result in a large β factor, however the non-radiative recombination rate does not impact on the fit considerably.[35] For the steady state solution of the above rate-equations, we set $dN/dt = 0$ and $dP/dt = 0$ to obtain.[34]

$$R_\text{p} = \frac{v_g g_{th} P}{\left(aP + \beta \frac{1}{\tau_\text{sp} v_g}\right)} \left(aP + \frac{1}{\tau_\text{sp} v_g}\right) \tag{3}$$

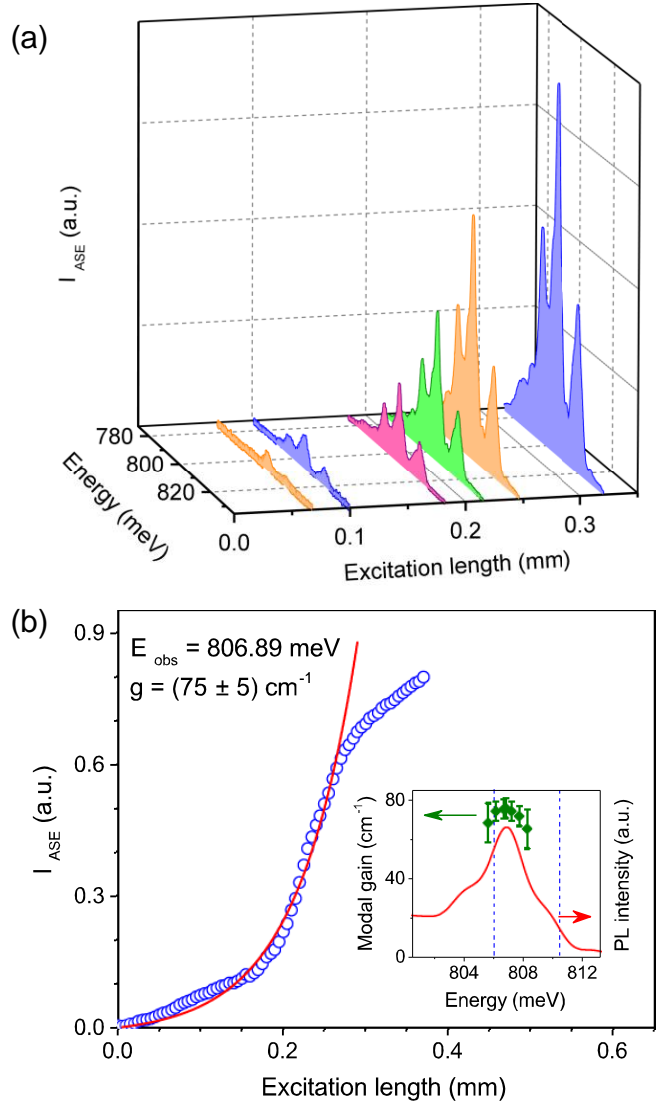

**Figure 3.** Optical amplification observation through the variable stripe length setup. (a) Edge emission PL spectra from a GaN epilayer under the above bandgap excitation, λ_exc = 351 nm, with different excitation lengths, and the pump intensity of 240 mJ.cm⁻². (b) PL intensity of the edge-emission for the PL line at 806.89 meV as a function of excitation length. The open circles are experimental data, and the solid line is the fitted curve using the one-dimensional amplification model. Inset provides the spectral dependence of the modal gain for output intensity around 806.89 meV.

We have fitted the experimental L-L curve to Equation 3 by first finding the material constant, $a$, that matches the measured threshold pump fluence at which a nonlinear jump of the output intensity was observed. Then the spontaneous emission factor, β, was varied until the best fit to the experimental data points was obtained. The lasing threshold for the excitation length of 0.20 mm was ~110 mJ.cm$^{-2}$ of the pump fluence, which corresponds to the spontaneous emission coupling factor, β = 0.3 (Fig. 2a). For comparison, L-L curves for several $β$ values are included, clearly showing the condition of the lasing threshold. When the pump intensity is higher than the threshold for ASE, most of Er ions in the excited state are stimulated to generate photons into waveguide modes, resulting in a large amount of photons emitted from the edge of the waveguide.[32, 36-38] A net optical amplification has been achieved. Correspondingly, for the fixed optical excitation length of 0.20 mm, the FWHM of the peak at 806.89 meV becomes narrow, and is about 1.65 ± 0.35 meV when the fluence of the optical pump is above 110 mJ.cm$^{-2}$ (Fig. 2b). Note that the lasing threshold in the GaN:Er epilayer is a few times higher compared to those from Er-doped GaN nanolayers in multiple quantum well GaN/AlN structures.[17, 39] In these structures, the multiple quantum wells provide an enhancement of the quantum efficiency for the infrared (1.5-μm) emission band through the strain engineering, and carrier/exciton quantum confinement effect.[17, 39]

To explore the gain coefficient from the peaks of PL emission, we have used the well-known VSL method. The active GaN:Er epilayer was excited optically in a stripe geometry by the UV laser providing the above bandgap excitation. In the VSL method, the spontaneous emission, $I_{SPONT}$, increase significantly when it passes the excitation volume of the sample within the waveguide. As a result of the increasing, the ASE signal, $I_{ASE}$, becomes visible and depends on the excitation length from the edge of the sample. Figure 3 presents the ASE intensity as a function of the excitation length. Using a one-dimensional model for the amplification, the modal gain, $g$, can be obtained from the excitation length, $l$, and the intensity, $I_{ASE}$:[29]

$$I_{ASE}(l,\lambda) = \frac{A \cdot I_{SPONT}}{g}\left(e^{g \cdot l} - 1\right) \qquad (4)$$

where $A$ is a constant related to the cross-section for the spontaneous emission, and the modal gain, $g$, is the net optical gain coefficient defined the gain minus losses of the material.

The increase of the excitation length obviously corresponds to the increase of optical pump excitation. When the optical pump excitation is high enough to achieve an optical amplification, the spontaneous emission is amplified exponentially by simulated emission as they travel through the active medium of the waveguide, resulting in an exponential increase in emission. Consequently, with the increase of the excitation length, we have observed an appearance of the dominant sharp peaks at 797.71 and 806.89 meV over the spontaneous emission and the emission intensity behavior changes from linear to exponential function at the sharp peaks (Fig. 3a). This is a direct signature of optical gain from Er optical centers in GaN epilayer around 1.5 μm at RT. We focus to analyze the strongest PL line at 806.89 meV. Under the pump fluence of 240 mJ.cm$^{-2}$, the emission intensity as a function of the excitation length for the sharp spectral peak exhibits a clear super-linear behavior. The red solid curve in Fig. 3b represents the best fitting curve based on Equation 4 with the net optical gain of ~75 ± 5 cm$^{-1}$. At a long excitation length (≥ 0.28 mm), the amplifier spontaneous emission shows a saturation behavior caused by the depletion of excited Er$^{3+}$ centers as the intense light travels through the waveguide. We also have performed the spectral dependence of modal gain in this narrow simulated emission bandwidth (Fig. 3b, inset), from which we have obtained the model gain at the PL line of 75 ± 10 cm$^{-1}$. The high model gain provides us a possibility of the near infrared lasing under electrical pumping by incorporating GaN epilayers into GaN light emitting structures.

## III. CONCLUSION

In summary, we have investigated the light amplification and stimulated emission of Er optical centers in GaN epilayers (bulk GaN crystal) prepared by MOCVD under the above bandgap excitation. Lasing action has been observed at 1.5 $\mu$m at RT. The observation is accompanied by the stimulated threshold, spectral linewidth narrowing, and strong modal gain in this material. The high modal gain indicates that it is feasible to realize electrically pumped lasing in the GaN:Er epilayers grown by MOCVD technique for high energy and high power semiconductor laser. The realization of near infrared lasers based on GaN epilayers would provide new opportunities in optoelectronics integration with high power electronics, photonics applications.

## ACKNOWLEDGMENTS

N.Q.V. acknowledges the support from NSF (ECCS-1358564). The materials growth effort at TTU as was supported by JTO/ARO (W911NF-12-1-0330).

## REFERENCES

[1] A. J. Kenyon, Erbium in silicon, Semicond. Sci. Tech. **20**, R65 (2005).
[2] G. H. Dieke, Spectra and energy levels of rare earth ions in crystals, Wiley, New York, (1968).
[3] A. J. Steckl, J. C. Heikenfeld, D. S. Lee, M. J. Garter, C. C. Baker, Y. Q. Wang, and R. Jones, Rare-earth-doped GaN: Growth, properties, and fabrication of electroluminescent devices, IEEE J. Sel. Top. Quant. **8**, 749 (2002).
[4] A. J. Steckl, J. H. Park, and J. M. Zavada, Prospects for rare earth doped GaN lasers on Si, Mater. Today. **10**, 20 (2007).
[5] A. J. Kenyon, Recent developments in rare-earth doped materials for optoelectronics, Prog. Quant. Electron. **26**, 225 (2002).
[6] N. Q. Vinh, N. N. Ha, and T. Gregorkiewicz, Photonic Properties of Er-Doped Crystalline Silicon, P. IEEE **97**, 1269 (2009).
[7] W. J. Miniscalco, Erbium-Doped Glasses for Fiber Amplifiers at 1500-nm, J. Lightwave. Technol. **9**, 234 (1991).
[8] M. A. Lourenco, R. M. Gwilliam, and K. P. Homewood, Extraordinary optical gain from silicon implanted with erbium, Appl. Phys. Lett. **91**, 141122 (2007).
[9] H. Przybylinska, W. Jantsch, Y. SuprunBelevitch, M. Stepikhova, L. Palmetshofer, G. Hendorfer, A. Kozanecki, R. J. Wilson, and B. J. Sealy, Optically active erbium centers in silicon, Phys. Rev. B **54**, 2532 (1996).
[10] N. Q. Vinh, H. Przybylinska, Z. F. Krasil'nik, and T. Gregorkiewicz, Optical properties of a single type of optically active center in Si/Si : Er nanostructures, Phys. Rev. B **70**, 115332 (2004).
[11] M. Wojdak, M. Klik, M. Forcales, O. B. Gusev, T. Gregorkiewicz, D. Pacifici, G. Franzo, F. Priolo, and F. Iacona, Sensitization of Er luminescence by Si nanoclusters, Phys. Rev. B **69**, 233315 (2004).
[12] H. S. Han, S. Y. Seo, and J. H. Shin, Optical gain at 1.54 mu m in erbium-doped silicon nanocluster sensitized waveguide, Appl. Phys. Lett. **79**, 4568 (2001).
[13] N. Daldosso, M. Luppi, S. Ossicini, E. Degoli, R. Magri, G. Dalba, P. Fornasini, R. Grisenti, F. Rocca, L. Pavesi, S. Boninelli, F. Priolo, C. Spinella, and F. Iacona, Role of the interface region on the optoelectronic properties of silicon nanocrystals embedded in $SiO_2$, Phys. Rev. B **68**, 085327 (2003).

[14] C. J. Oton, W. H. Loh, and A. J. Kenyon, $Er^{3+}$ excited state absorption and the low fraction of nanocluster-excitable $Er^{3+}$ in $SiO_x$, Appl. Phys. Lett. **89**, 031116 (2006).
[15] P. N. Favennec, H. Lharidon, M. Salvi, D. Moutonnet, and Y. Leguillou, Luminescence of Erbium Implanted in Various Semiconductors - IV-Materials, III-V-Materials and II-VI Materials, Electron. Lett. **25**, 718 (1989).
[16] F. Elmasry, S. Okubo, H. Ohta, and Y. Fujiwara, Electron spin resonance study of Er-concentration effect in GaAs; Er, O containing charge carriers, J. Appl. Phys. **115**, 193904 (2014).
[17] V. X. Ho, T. M. Al Tahtamouni, H. X. Jiang, J. Y. Lin, J. M. Zavada, and N. Q. Vinh, Room-Temperature Lasing Action in GaN Quantum Wells in the Infrared 1.5 mu m Region, ACS Photonics **5**, 1303 (2018).
[18] I. W. Feng, X. K. Cao, J. Li, J. Y. Lin, H. X. Jiang, N. Sawaki, Y. Honda, T. Tanikawa, and J. M. Zavada, Photonic properties of erbium doped InGaN alloys grown on Si (001) substrates, Appl. Phys. Lett. **98**, 081102 (2011).
[19] C. Ugolini, N. Nepal, J. Y. Lin, H. X. Jiang, and J. M. Zavada, Erbium-doped GaN epilayers synthesized by metal-organic chemical vapor deposition, Appl. Phys. Lett. **89**, 151903 (2006).
[20] V. X. Ho, T. V. Dao, H. X. Jiang, J. Y. Lin, J. M. Zavada, S. A. McGill, and N. Q. Vinh, Photoluminescence quantum efficiency of Er optical centers in GaN epilayers, Sci. Rep. **7**, 39997 (2017).
[21] H. Shibata, Y. Waseda, H. Ohta, K. Kiyomi, K. Shimoyama, K. Fujito, H. Nagaoka, Y. Kagamitani, R. Simura, and T. Fukuda, High thermal conductivity of gallium nitride (GaN) crystals grown by HVPE process, Mater. Trans. **48**, 2782 (2007).
[22] C. Roder, S. Einfeldt, S. Figge, and D. Hommel, Temperature dependence of the thermal expansion of GaN, Phys. Rev. B **72**, 085218 (2005).
[23] R.R. Reeber, and K. Wang, Lattice parameters and thermal expansion of GaN, J. Mater. Res. **15**, 40 (2000).
[24] Group IV Elements, IV-IV and III-V Compounds. Part b - Electronic, Transport, Optical and Other Properties, Springer, Berlin, Heidelberg (2002).
[25] J. H. Kim, D. Seong, G. H. Ihm, and C. Rhee, Measurement of the thermal conductivity of Si and GaAs wafers using the photothermal displacement technique, Int. J. Thermophys. **19**, 281 (1998).
[26] A. I. Zagumennyi, G. B. Lutts, P. A. Popov, N. N. Sirota, and I. A. Shcherbakov, The Thermal Conductivity of YAG and YSAG Laser Crystals, Laser Phys. **3**, 1064 (1993).
[27] H. Furuse, R. Yasuhara, and K. Hiraga, Thermo-optic properties of ceramic YAG at high temperatures, Opt. Mater. Express **4**, 1794 (2014).
[28] R. Dahal, J. Y. Lin, H. X. Jiang, and J. M. Zavada, Er-Doped GaN and $In_xGa_{1-x}N$ for Optical Communications, Top. Appl. Phys. **124**, 115 (2010).
[29] K. L. Shaklee, and R. F. Leheny, Direct Determination of Optical Gain in Semiconductor Crystals, Appl. Phys. Lett. **18**, 475 (1971).
[30] D. K. George, M. D. Hawkins, M. McLaren, H. X. Jiang, J. Y. Lin, J. M. Zavada, and N. Q. Vinh, Excitation mechanisms of Er optical centers in GaN epilayers, Appl. Phys. Lett. **107**, 171105 (2015).
[31] C. Ugolini, N. Nepal, J. Y. Lin, H. X. Jiang, and J. M. Zavada, Excitation dynamics of the 1.54 um emission in Er doped GaN synthesized by metal organic chemical vapor deposition, Appl. Phys. Lett. **90**, 051110 (2007).
[32] J. C. Johnson, H. J. Choi, K. P. Knutsen, R. D. Schaller, P. D. Yang, and R. J. Saykally, Single gallium nitride nanowire lasers, Nat. Mater. **1** 106 (2002).
[33] M. D. McGehee, and A. J. Heeger, Semiconducting (conjugated) polymers as materials for solid-state lasers, Adv. Mater. **12**, 1655 (2000).

[34] S. Strauf, K. Hennessy, M. T. Rakher, Y. S. Choi, A. Badolato, L. C. Andreani, E. L. Hu, P. M. Petroff, and D. Bouwmeester, Self-tuned quantum dot gain in photonic crystal lasers, Phys. Rev. Lett. **96**, 127404 (2006).
[35] G. Bjork, A. Karlsson, and Y. Yamamoto, Definition of a Laser Threshold, Phys. Rev. A **50**, 1675 (1994).
[36] N. Tessler, Lasers based on semiconducting organic materials, Adv. Mater. **11**, 363 (1999).
[37] L.A. Coldren, S.W. Corzine, Diode Lasers and Photonic Integrated Circuits, Wiley, New York, (1995).
[38] A. E. Siegman, Lasers, University Science Books, Susalito, (1986).
[39] T. M. Al Tahtamouni, M. Stachowicz, J. Li, J. Y. Lin, and H. X. Jiang, Dramatic enhancement of 1.54 um emission in Er doped GaN quantum well structures, Appl. Phys. Lett. **106**, 121106 (2015).